\documentstyle[prl,multicol,aps,epsfig]{revtex}

\draft

\begin{document}
\title{Strong enhancement of drag and dissipation at the weak- to strong- coupling
phase transition in a bi-layer system at a total Landau level filling $\nu=1 
$}
\date{\today }
\author{Ady Stern$^{(1)}$ and B.I. Halperin$^{(2)}$}
\address{(1) Department of Condensed Matter Physics, Weizmann Institute of Science,\\
Rehovot 76100, Israel\\
(2) Physics Department, Harvard University, Cambridge, Mass. 02138}
\maketitle

\begin{abstract}
We consider a bi-layer electronic system at a total Landau level filling
factor $\nu =1$, and focus on the transition from the regime of weak
inter-layer coupling to that of the strongly coupled $(1,1,1)$ phase (or
''quantum Hall ferromagnet''). Making the assumption that in the transition
region the system is made of $\ $puddles of the $(1,1,1)$ phase embedded in
a bulk of the weakly coupled state, we show that the transition is
accompanied by a strong increase in longitudinal Coulomb drag, that reaches
a maximum of approximately $h/2e^{2}$. In that regime the longitudinal drag
is increased with decreasing temperature.
\end{abstract}

\begin{multicols}{2}
\bigskip Bi-layer electronic systems at a total Landau level filling factor $%
\nu =1$ exhibit a very rich phase diagram with striking physical properties 
\cite{smg-review,perspectivesbook}. A useful parameter for exploring this
phase diagram is the ratio between the inter-layer separation, $d$ and the
magnetic length, $l$. Experimental probes of the bi-layer system are
measurements of the symmetric resistivity matrix $\rho ^{s}$ (the
longitudinal and Hall resistivities to the flow of equal current in both
layers), measurements of the trans-resistivity $\rho ^{D}$ matrix (often
called the drag resistivity), in which current flows in one (''active'')
layer and voltage is measured on the other (''passive'') one, and
measurements of the tunneling differential conductance $dI/dV$.

Systems of large layer separation ($d/l\simeq 2-4$) have been thoroughly
studied experimentally. Their linear response may be described as that of
two weakly coupled resistors in parallel, each at $\nu =1/2$. In particular,
a perturbative analysis of the inter-layer coupling accounts
for the suppression of inter-layer tunneling at low voltages \cite{he} and
the strong enhancement of longitudinal Coulomb drag as compared to its value
at zero magnetic field\cite{ussishkin,sakhi,lilly,kim}. In this analysis the
system is regarded as composed of two weakly coupled composite fermion
metals.

In contrast, the limit of small $d/l$ is believed to be
adequately described in terms of strong inter-layer correlations.
Theoretical models of this limit describe it in terms of a $(1,1,1)$ state,
a quantum Hall ferro-magnet, a condensate of composite bosons or a
condensate of excitons\cite{smg-review}. Recent experiments (where $%
d/l\lesssim 2)$ probe this limit and give strong support to these models.
The observed phenomena, including a quantum Hall state in the symmetric
channel\cite{Kellogg}, a very large zero voltage peak in the tunneling
differential conductance\cite{jpetunnel1}, a splitting of this peak by the
application of a parallel magnetic field\cite{jpegoldstone} and a
quantization of the Hall drag resistivity \cite{Kellogg} are all
consequences of these models\cite{ady2PRL,leonleoPRL,fogler}.

Very little is known about the transition between the two limits. Numerical
evidence suggests a first order zero temperature phase transition \cite
{Schlieman}. However, experimental observations show that as the ratio $d/l$
is decreased, both the zero voltage tunneling peak and the Hall drag
resistivity develop very gradually \cite{jpetunnel1,Kellogg}. Moreover,
preliminary measurements of longitudinal drag near the transition show a
non-monotonic behavior. As $d/l$ is reduced towards the transition, the
longitudinal drag grows dramatically. Then, when the transition region is
reached, the longitudinal drag decreases and practically vanishes when the $%
\nu =1$ quantum Hall state develops \cite{jpe-priv}.

In this paper we postulate that in the transition region the system is composed of puddles of the $%
(1,1,1)$ phase occupying a fraction $f$ of the sample, and puddles of the
weakly coupled phase occupying the rest. As the ratio $d/l$ is decreased,
the fraction $f$ increases. When $f$ reaches a critical value $f_{c}$ the $%
(1,1,1)$ phase percolates. Here we assume $f_{c}=1/2$, which is the case for
a statistically symmetric distribution of phases. Since our analysis focuses
on the transition region, all expressions given below refer to the range $%
0\leq f\leq 1/2$. Furthermore, we assume that the puddles of each phase are
large enough such that a local conductivity may be defined for each puddle.

We explore the consequences of this model regarding drag in the transition
region. In particular, we find that this model naturally explains the
non-monotonic behavior of longitudinal drag in that region. Although both
the weakly coupled phase and the $(1,1,1)$ phase are characterized by very
little dissipation, we find that for a mixed system rather close to $f=1/2$,
a flow of current in one layer is accompanied by very large dissipation,
with a dissipative resistivity peaking at $h/2e^{2}.$

The coexistence of the two phases in the transition region stems from
fluctuations of the electron density. If smooth enough, these fluctuations
allow for phase separation. For \ a fluctuation $\delta \nu $ in filling
factor and for a circular puddle, the fluctuation in density translates to a
fluctuation in the number of electrons in the puddle when the puddle is much
larger than $l_{H}\sqrt{\frac{8}{\delta \nu }.}$ Assuming a typical
fluctuations $\delta \nu _{typ}=0.04$ this condition requires the puddles to
be larger than about $8000$\AA . To qualitatively understand the relation of
the fraction $f$ to the ratio $d/l$ and to $\delta \nu _{typ}$ we consider
the phase diagram on a plane whose axes are $d/l$ and $\nu .$ For $%
\nu =1$ the $(1,1,1)$ phase occupies the range $0<\frac{d}{l}<\left( 
\frac{d}{l}\right) _{c}$ where the critical value $\left( \frac{d}{l}\right)
_{c}$ is theoretically estimated to be $\approx 2$ and experimentally
appears to be $\approx 1.8.$ The $(1,1,1)$ phase is very sensitive to
deviations of the total filling factor from $\nu =1,$ which introduce
topological defects into its order parameter \cite
{smg-review,perspectivesbook}. Thus, for each value of $d/l$ there is a
limited range of filling factors around $\nu =1$ for which the system can
still be described in terms of a perturbed $(1,1,1)$ state. The width of
that region, $2\Delta \nu ,$ is zero at $\left( \frac{d}{l}\right) _{c}$ and
grows with decreasing $d/l.$ It may be extracted from the width of the
plateau of the observed $\rho _{xy}^{D}.$ The experiment shows $\Delta \nu
\approx 0.04$ for $d/l=1.6$ and $\Delta \nu \approx 0.03$ for $d/l=1.66$ 
\cite{Kellogg}. Crudely, we may view the distribution of filling factors in
the sample as a uniform distribution with a width $2\delta \nu _{typ}.$
Within this picture, the fraction $f=\min \left( \frac{\Delta \nu }{\delta
\nu _{typ}},1\right) .$ It is hard to make this relation more explicit,
since neither $\delta \nu _{typ}$ nor the dependence of $\Delta \nu $ on $d/l
$ are well characterized. 

In the absence of inter-layer tunneling, the linear response of a bi-layer
system is described by a $4\times 4$ matrix, with the four rows being $%
(1,x),(1,y),(2,x),(2,y)$ (here the number indicates the layer and the Latin
letter indicates a Cartesian direction). When the two layers are identical,
the linear response matrices are block-diagonal in a basis of symmetric and
anti-symmetric states: a symmetric (anti-symmetric) current generates a
purely symmetric (anti-symmetric) electric field. We denote the $2\times 2$
resistivity matrices corresponding to symmetric and anti-symmetric currents
by $\rho ^{s},\rho ^{a},$ respectively$.$ In a drag measurement current is
applied in one layer while the voltage is measured in the other layer,
yielding a $2\times 2$ drag resistivity matrix, $\rho ^{D}=\frac{1}{2}\left(
\rho ^{a}-\rho ^{s}\right) $.

For a $\nu =1$ bi-layer system in the limit of large separation ($%
d/l\rightarrow \infty $, to be denoted by the subscript $\infty $) the
layers are only weakly coupled, and $\rho ^{D}$ is much smaller than both $%
\rho ^{a},\rho ^{s}.$ Thus, we may approximate (all resistivities are
expressed in units of $h/e^{2}$)$:$ 
\begin{equation}
\rho _{\infty }^{a}=\rho _{\infty }^{s}=\left( 
\begin{array}{cc}
\varepsilon  & 2 \\ 
-2 & \varepsilon 
\end{array}
\right)   \label{rhoinfinity}
\end{equation}
The value of $\varepsilon $ depends on the quality of the sample, and is
usually of the order of $0.05-0.3$ \cite{jpe-priv}. At $T\lesssim 1^{0}K$
the experimentally measured values of $\rho ^{D}$ are smaller than $%
\varepsilon $ by a factor of at least $50.$ Within a composite fermion model
of the two weakly coupled layers, $\varepsilon =\left( k_{F}l_{tr}\right)
^{-1}$, (where $k_{F}\equiv 4\pi n$ is the Fermi wave-vector and $n$ is the
electronic density in each layer). For a local conductivity to be well
defined in the weakly coupled phase a typical puddle size should be much
larger than $l_{tr}=\left( \varepsilon \sqrt{4\pi n}\right) ^{-1},$ which
for typical values is about $2000$\AA .

The limit of small separation, where the system is in the $(1,1,1)$ phase, is denoted by the subscript $0$. For symmetric currents it
is a quantum Hall state, while for anti-symmetric currents it is a
superfluid, i.e., 
\begin{equation}
\rho _{0}^{s}=\left( 
\begin{array}{cc}
0 & 2 \\ 
-2 & 0
\end{array}
\right) \text{ \ \ and \ \ \ }\rho _{0}^{a}=0.  \label{rho111}
\end{equation}

In the composite system, puddles of the $(1,1,1)$ phase are embedded in a
bulk of the $d=\infty $ phase. 
 The motion of current into and out of the puddles may involve a boundary resistance that is inversely proportional to the puddles' perimeter. We assume this size large enough for boundary resistance to be negligible. Lacking a microscopic model for this resistance, however, we cannot quantify this condition. 
In the absence of inter-layer tunneling both
symmetric and anti-symmetric components of the currents and electric fields
satisfy Kirchoff's equations ${\bf \nabla} \cdot {\bf j}^{s(a)}={\bf \nabla} \times
{\bf E}^{s(a)}=0. $  
The analysis of the symmetric resistivity is rather simple.
Since both phases have $\rho _{xy}^{s}=2$, the composite system has $\rho
_{xy}^{s}=2$ as well. The solutions for ${\bf j}^{s},{\bf E}^{s}$
can be found by first solving for the current and electric field $%
{\bf j}_{0,}^{s}{\bf E}_{0}^{s}$ with $\rho _{xy}^{s}$ set to zero, and then identifying 
${\bf j}^{s}={\bf j}_{0}^{s}$ and ${\bf E}^{s}={\bf E}_{0}^{s}+\frac{2h}{e^{2}}\hat{z}\times {\bf j}^{s}.$
The value of the longitudinal resistivity $\rho _{xx}^{s}$ then varies from $%
\varepsilon $ for $f=0$ to zero for $f=1/2.$ Within the (rough) effective
medium approximation, $\rho _{xx}^{s}\simeq \varepsilon (1-2f)$ \cite
{bergman}. For the purpose of calculating the symmetric longitudinal {\it %
resistivity}, then, our composite system maps onto super-conducting inclusions in a two dimensional conductor.

For the analysis of the anti-symmetric resistivity, we note that
since $\rho _{0}^{a}=0$, its inverse is not uniquely defined. In particular,
we could write it as $\sigma _{0}^{a}=\left( 
\begin{array}{cc}
\infty  & -2/(4+\varepsilon ^{2}) \\ 
2/(4+\varepsilon ^{2}) & \infty 
\end{array}
\right) .$ This choice makes $\sigma_{xy}^{a}$ of both phases equal, and
allows an analysis similar to that of the symmetric case: since the Hall
conductivity is uniform, the solutions to Kirchoff's equations, $j^{a},E^{a}$, are related to the current and electric field $j_{0,}^{a}E_{0}^{a}$ in the
absence of a magnetic field ($\sigma _{xy}^{a}=0)$, by $j^{a}=j_{0}^{a}-%
\frac{2e^{2}}{(4+\varepsilon ^{2})h}\hat{z}\times E_{0}^{a}$ and $%
E^{a}=E_{0}^{a}.$ The longitudinal conductivity $\sigma _{xx}^{a}$ is then
unaffected by the magnetic field. Again, the analogy to 
super-conducting inclusions embedded in a conducting medium can be used, but
this time for the calculation of the anti-symmetric {\it conductivity}.
Thus, $\sigma _{xx}^{a}$ varies strongly with $f.$ It is $\frac{\varepsilon 
}{4+\varepsilon ^{2}}$ for $f=0$ and becomes infinite at $f\geq 1/2$. Within the
effective medium approximation $\sigma _{xx}^{a}\simeq \frac{%
\varepsilon }{4+\varepsilon ^{2}}\frac{1}{1-2f}$ below the percolation
threshold. With the Hall conductivity being $\approx 1/2$ and the
longitudinal conductivity changing from $\varepsilon \ll 1/2$ to infinity,
the longitudinal resistivity is non-monotonic with respect to $f$. It is  $\rho _{xx}^{a}=%
\frac{\varepsilon \left( 4+\varepsilon ^{2}\right) \left( 1-2f\right) }{%
\varepsilon ^{2}+4\left( 1-2f\right) ^{2}}.$ Consequently, the longitudinal
drag resistivity is 
\begin{equation}
\rho _{xx}^{D}=\frac{8\varepsilon f\left( 1-f\right) \left( 1-2f\right) }{%
\varepsilon ^{2}+4\left( 1-2f\right) ^{2}}  \label{dragformula}
\end{equation}
This is an interesting expression: as $f$ increases from zero, the longitudinal drag increases from zero,
reaching a maximum at a fraction $f^{\ast }$ below the percolation
threshold, and vanishing back to zero at percolation. For small $\varepsilon 
$, the maximum of the longitudinal drag takes place at $f^{\ast }=\frac{1}{2}%
-\frac{\varepsilon }{4},$ where $\rho _{xx}^{D}\approx \frac{h}{2e^{2}}.$
The drag resistivity $\rho _{xx}^{D}$ is plotted as a function of $f$ in
Fig. [1], for $\varepsilon =0.1$ and $\varepsilon =0.3$. For small $\varepsilon$, improvements to the
effective medium approximation may modify the $f$-dependence in (\ref{dragformula}), but would not vary the maximum value.


{\setlength{\unitlength}{1cm}
\begin{figure}
\centerline{\begin{picture}(4,4.5)
   \put(-0.8,0){\psfig{width=6cm,figure=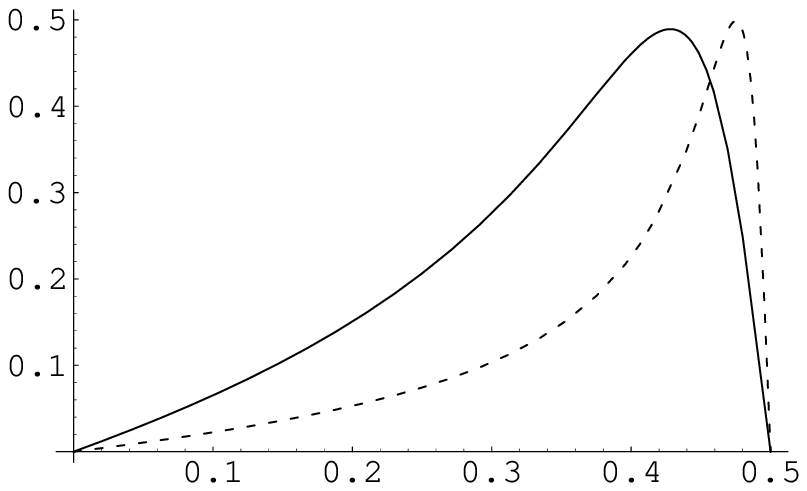}}
   \put(5.0,-0.5){\makebox(0,0)[b]{${ f}$}}
   \put(-1.2,3.0){\makebox(0,0)[b]{$\rho_{xx}^D$}}
\end{picture}}
\vspace*{1.2cm}
\caption{ The longitudinal drag resistivity $\rho_{xx}^D$ as a function of $f$
for $\varepsilon=0.1$ (dashed line) and $\varepsilon=0.3$ (solid line).}

\end{figure}}

Several comments should now be made regarding this behavior of the
longitudinal drag. First, the peak in $\rho _{xx}^{D}$ is limited to a
rather narrow range of $f$: for small $\varepsilon $, the longitudinal drag
resistivity is below a quarter of its peak value (i.e., $<h/8e^{2}$) at $f=%
\frac{1}{2}-2\varepsilon ,$ below half of its peak value ($<h/4e^{2}$) when $%
f=\frac{1}{2}-\varepsilon ,$ reaches its maximum at $f=f^{\ast }$ and
vanishes at $f=\frac{1}{2}.$ Consequently, the peak would be narrow also on
the $d/l$ axis, as Hall drag measurements indicate that $f$ changes from
zero to $1/2$ over a range of $d/l$ of about $0.15.$ Second, the strong
longitudinal drag at the transition region is accompanied by local
circulating currents in the passive layer. The current density in the
passive layer is $\frac{1}{2}\left( j^{a}-j^{s}\right) $. Since no current
flows into and out of the passive layer, this layer carries no longitudinal
current, and thus the longitudinal parts of the symmetric and anti-symmetric
currents are identical. However, as explained above, the transverse parts of
the current differ: $\frac{1}{2}\nabla \times \left( j^{a}-j^{s}\right) =$ $%
\frac{e^{2}}{(4+\varepsilon ^{2})h}\nabla \cdot E^{a}$. The anti-symmetric
electric field vanishes inside the $(1,1,1)$ puddles, where the conductivity
is infinite, and is generally non-zero in the bulk. Thus, interfaces between 
$(1,1,1)$ puddles and the bulk are generally accompanied by circulating
currents. And third, the increase in drag goes hand in hand with a similar
increase in the resistivity of one layer, when measured with no current
driven in the second layer. This resistivity, which is $\frac{1}{2}\left(
\rho _{xx}^{a}+\rho _{xx}^{s}\right) ,$ has a maximum close to $\frac{h}{%
2e^{2}}$ similar to that of the longitudinal drag. Unlike drag, however,
this resistivity is a measure of dissipation, indicating that dissipation is
strongly enhanced very close to the transition.

Further insight into drag in the composite system can be gained by employing
a semi-circle law, of the type derived, e.g., by Dykhne and Ruzin \cite
{dykhne} for a two dimensional composite system made of two components. By
this law, the longitudinal and Hall components of the macroscopic
resistivity tensor of a system composed of a mixture of two phases satisfy a
semi-circle relation 
\begin{equation}
\left( \rho _{xy}-\bar{\rho}_{xy}\right) ^{2}+\rho _{xx}^{2}=\bar{\rho}%
_{xx}^{2}
\end{equation}
The values of $\bar{\rho}_{xx},\bar{\rho}_{xy}$ are determined by the
longitudinal and Hall resistivities of the two phases: 
\begin{eqnarray}
\bar{\rho}_{xy} &=&\frac{1}{2}\frac{\det \rho ^{(2)}-\det \rho ^{(1)}}{\rho
_{xy}^{(2)}-\rho _{xy}^{(1)}}  \nonumber \\
\bar{\rho}_{xx} &=&\sqrt{\bar{\rho}_{xy}^{2}+\frac{\det \rho ^{(2)}\rho
_{xy}^{(1)}-\det \rho ^{(1)}\rho _{xy}^{(2)}}{\rho _{xy}^{(1)}-\rho
_{xy}^{(2)}}}  \label{semi-circ-general}
\end{eqnarray}
with $\rho ^{(i)}$ being the resistivity matrix of the $i^{\prime }$th
phase. For a bi-layer system of two identical layers, this law can be
independently applied to $\rho ^{s}$ and $\rho ^{a}$. For the symmetric
tensor, Eqs. (\ref{rhoinfinity})-(\ref{semi-circ-general}) yield a
semi-circle of infinite radius. As the fraction $f$ is turned from zero to
one, the system goes over an infinitesimal part of the infinite semi-circle,
with the Hall component remaining constant and the longitudinal component
varying monotonically from $\varepsilon $ to zero. For the anti-symmetric
resistivity Eqs. (\ref{rhoinfinity})-(\ref{semi-circ-general}) yield $\bar{%
\rho}_{xy}^{a}=\bar{\rho}_{xx}^{a}=\left( 1+\varepsilon ^{2}/4\right) .$ As $%
f$ increases from zero to one, the anti-symmetric resistivity tensor goes
over most of the semi-circle, and its longitudinal component varies
non-monotonically. As a consequence, the symmetric and
drag resistivities follow the relation: 
\begin{equation}
\left( \rho _{xy}^{D}+\frac{1}{2}(1-\frac{\varepsilon ^{2}}{4})\right)
^{2}+\left( \rho _{xx}^{D}+\frac{1}{2}\rho _{xx}^{S}\right) ^{2}=\frac{1}{4}%
\left( 1+\frac{\varepsilon ^{2}}{4}\right) ^{2}
\end{equation}
To lowest order in $\varepsilon ,$ this is a semi-circle relation for the
drag resistivity tensor, 
\begin{equation}
\left( \rho _{xy}^{D}+\frac{1}{2}\right) ^{2}+\left( \rho _{xx}^{D}\right)
^{2}=\frac{1}{4}.
\end{equation}
For $f=0$ both components of $\rho ^{D}$ are zero (more precisely, much
smaller than $\varepsilon $). For $f\geq 1/2,$ the longitudinal drag vanishes and the Hall drag assumes
the quantized value $\rho _{xy}^{D}=-1.$ In between these two extremes a
negative $\rho _{xy}^{D}$ develops monotonically, while $\rho _{xx}^{D}$
develops non-monotonically from zero to $1/2$ and back to zero.

We now turn to discuss the dependence of the longitudinal drag on the total
filling factor and on temperature. The phase diagram of
the $\nu =1$ bi-layer with respect to these parameters is not quantitatively
known, but we can still make some 
qualitative predictions. First, it is plausible that the fraction $f$
decreases as temperature is increased, until it vanishes at a
Kosterlitz-Thouless (K-T) phase transition. Consequently, as long as $f<f^{\ast }$ (i.e., throughout most of
the transition region), a composite system would show a {\it longitudinal
drag that increases with decreasing temperature}
in sharp contrast to the behavior of drag deep in the weak coupling and $%
(1,1,1)$ regimes. Indeed, preliminary measurements show such a trend \cite
{jpe-priv}. Second, since the $(1,1,1)$ state is very sensitive to the total
bi-layer filling factor, it is plausible that the fraction $f$ decreases
when the total filling factor is varied away from $\nu =1.$ 
Thus, for values of $d/l$ where $f<f^{\ast }$ at $\nu =1,$ the drag
resistivity $\rho _{xx}^{D}$ would be maximal at $\nu =1.$ In contrast, for
values of $d/l$ where $f>f^{\ast }$ at $\nu =1,$ the drag resistivity $\rho
_{xx}^{D}$ would have a local minimum at $\nu =1$, surrounded by two local
maxima around $\nu =1$. These two maxima are not necessarily symmetric
around $\nu =1$. To lowest order in $\varepsilon $ the value of the longitudinal drag at
these maxima  is $1/2\nu ^{\ast }$ (where $\nu ^{\ast }$ is the filling
factor at which $f=f^{\ast }$).

Our two-phase model of the transition region has also implications regarding
inter-layer tunneling, which we briefly comment on. The
tunneling characteristics of the two phases are strikingly different: in the
weak coupling phase low voltage tunneling is strongly suppressed, while in
an ideal $(1,1,1)$ phase an infinite zero voltage peak is expected in the
differential conductance $\frac{dI}{dV}.$ A sharp, but finite, peak is
observed\cite{jpetunnel1}. A quantitative analysis of inter-layer tunneling requires a
microscopic understanding of the mechanism that cuts off the divergence
in $\frac{dI}{dV}|_{V=0}$ in the $(1,1,1)$ phase, and this understanding is
presently lacking (see Refs. \cite{ady2PRL}--\cite{fogler}
for a phenomenological discussion). We point out, however, that one should not expect the range of $%
f $ where dissipation and drag are strongly enhanced to coincide with the
evolution of  the tunneling peak. The measurements of tunneling deep in
the $(1,1,1)$ phase may be taken to indicate that $\frac{dI}{dV}|_{V=0}$  in that phase is finite, rather than infinite, at
least at non-zero temperature. If that is the case, and as long as tunneling
may be regarded as a perturbation, the tunneling $\frac{dI}{dV}|_{V=0}$  should be roughly proportional to $f$ even after the $(1,1,1)$
phase percolates, i.e., when $f>1/2.$ Then, as $d/l$ is decreased, the
tunneling peak continues evolving long after the longitudinal drag vanishes
and the Hall drag gets quantized. The position and width of the transition
region then appear different when inferred from tunneling or from drag
measurements.

Previous studies have raised the possibility of an intermediate phase
emerging in the transition between the weak coupling and $(1,1,1)$ phases.
Candidates for such a phase were described in terms of a
coupled Wigner crystal\cite{cote}, inter-layer pairing of composite fermions
and condenstaions of various types of composite bosons\cite{kimread}. We believe that some
of our predictions may serve as a litmus test distinguishing between the scenario presented here and the emergence of a third
intermediate phase. In particular, we refer to the semi-circle behavior of
the ant-symmetric resistivity and a saturation of the longitudinal drag
resistivity peak with decreasing temperature.

We thank J.P. Eisenstein and M. Kellogg for discussing their data
with us prior to publication. This work was supported by the Israel-US BSF,
the Israeli Science Foundation, a grant from the German-Israel DIP
corporation and NSF grant  DMR-9981283.

\end{multicols}

\end{document}